\documentclass{article}%
\usepackage{amsmath}
\usepackage{amsfonts}
\usepackage{amssymb}
\usepackage{graphicx}%
\providecommand{\U}[1]{\protect\rule{.1in}{.1in}}

\begin{document}

\title{Improving the Pe\~{n}a-Prieto \textquotedblleft KSD\textquotedblright\ procedure}
\author{Ricardo Maronna (rmaronna@retina.ar)\\University of La Plata and University of Buenos Aires}
\maketitle
\date{}

\begin{abstract}
Pe\~{n}a and Prieto (2007) proposed the\textquotedblleft Kurtosis plus
specific directions\textquotedblright\ (KSD) method for robust multivariate
location and scatter estimation and outlier detection. Maronna and Yohai
(2017) employed it it as an initial estimator for multivariate S- and
MM-estimators, and their simulations showed that KSD generally outperforms
initial estimators based on subsampling. However further simulations show that
KSD may become unstable and give wrong results in extreme situations when the
contamination rate is \textquotedblleft high\textquotedblright\ ($\geq0.2)$
and the ratio $n/p$ of cases to variables is \textquotedblleft
low\textquotedblright\ (%
$<$%
10). Two simple modifications of the procedure are proposed, which greatly
improve on the method's performance as an initial estimator, with only a small
increase in computational time.

\end{abstract}

\section{The problem}

Pe\~{n}a and Prieto (2007) developed an elaborate procedure for multivariate
outlier detection and robust estimation of multivariate location and scatter,
called \textquotedblleft Kurtosis plus Specific Directions\textquotedblright%
\ (henceforth KSD). Maronna and Yohai (2017) have employed it as a starting
estimator for robust multivariate estimators that are computed iteratively and
need reliable initial values, and their simulations show that the initial
values supplied by KSD generally yield better results than those based on
subsampling, both in statistical performance and in computing speed.

However, it was observed by the author that in certain "difficult" \ cases KSD
can be highly unstable and yield totally useless values. These cases occur
when the contamination rate $\varepsilon$ is "high" ($\geq0.2)$ and the ratio
$n/p$ of number of observations to number of variables is "low" (%
$<$%
10).

The problem was studied through simulated data with $p$ between 10 and 50.
Contaminated normal samples were generated as in (Maronna and Yohai 2017),
namely: the \textquotedblleft clean\textquotedblright\ observations
$\mathbf{x}_{i}\in R^{p}$ ($i=1,...,n)$ are generated as $\mathrm{N}%
_{p}\left(  \mathbf{0,I}\right)  .$ Call respectively $\varepsilon,$ $K$ and
$\gamma$ the contamination rate, the outlier size and the outliers' dispersion
and let $m=[n\varepsilon]$. Then the data are contaminated by changing
$x_{i1}$ to $\gamma x_{i1}+K\ $for $i=1,...,m.$

Since we are interested in the performance of KSD as a starting estimator,
rather than the behavior of KSD itself, we use the output of KSD to start the
iterations to compute Rocke's (1996) S-estimator of location and scatter
$\left(  \widehat{\mathbf{\mu}}\mathbf{,}\widehat{\mathbf{\Sigma}}\right)  .$
The matrix $\widehat{\mathbf{\Sigma}}$ is corrected to make it consistent at
the normal.

As an example, we generate $N=200$ samples with $p=30,$ $n=100$ and
contamination rate $\varepsilon=0,2.$ and two values of $\gamma:$ 0 and 0.5,
corresponding to concentrated and moderately dispersed outliers. For each
sample we compute Rocke's location vector $\widehat{\mathbf{\mu}}$ and scatter
matrix $\widehat{\mathbf{\Sigma}}$, and for each of them we compute its
Kullback-Leibler divergence:%
\[
D\left(  \widehat{\mathbf{\mu}}\right)  =\left\Vert \widehat{\mathbf{\mu}%
}\right\Vert ^{2},\ \ D\left(  \widehat{\mathbf{\Sigma}}\right)
=\mathrm{trace}\left(  \widehat{\mathbf{\Sigma}}\right)  -\log\det\left(
\widehat{\mathbf{\Sigma}}\right)  -p.
\]

In the following we concentrate on the scatter matrix $\widehat{\mathbf{\Sigma
}}$, which appears to be more affected than $\widehat{\mathbf{\mu}}$ by this
phenomenon. The next figure shows the ordered values of $D\left(
\widehat{\mathbf{\Sigma}}\right)  $ for the $N$ Monte Carlo replications,
corresponding to $K=13$ with $\gamma=0$ and $\gamma$=$0.5$.%

\begin{center}
\includegraphics[
trim=0.000000cm 0.000000cm 0.000000cm 0.744966cm,
height=9.1752cm,
width=13.1143cm
]%
{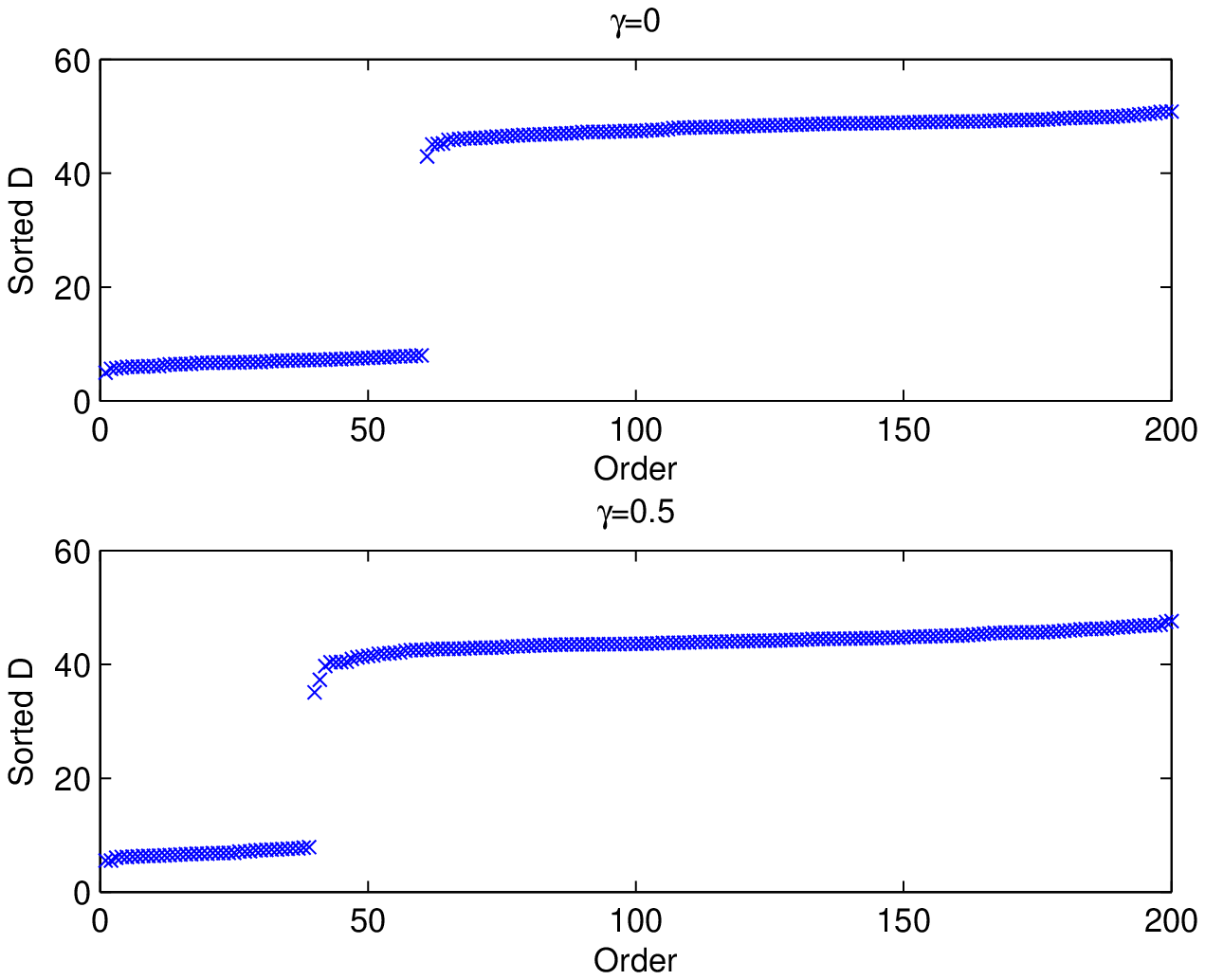}%
\\
Ordered values of $D\left(  \mathbf{\Sigma}\right)  $ of the Rocke estimator
starting from the old version of KSD for $K=13$%
\end{center}

It is seen that the estimator is rather unstable and may yield very high
values: for $\gamma=0$ 30\% of the values are below 8 and the rest are above
42; for $\gamma=0.5$ roughly 20\% of the values are below 8 and the rest are
above 35.

\section{The solution}

We first need to give a brief description of KSD. It consists of three stages.

The data are first normalized to zero means and identity covariance matrix.
Call $\mathbf{z}_{i},$ $i$=$1,...,n$ the normalized data and $\mathbf{Z}$ the
respective $n\times p$ matrix .

In stage I, a number $N_{\mathrm{Kurt}}$ of directions $\mathbf{u}_{k}$ is
derived, each of which yields a local maximum or minimum of the kurtosis of
the projections $\mathbf{Zu}_{k}\mathbf{.}$

In stage II a number $N_{\mathrm{SD}}$ of random \textquotedblleft specific
directions\textquotedblright\ $\mathbf{v}_{k}$ is computed, by a sort of
stratified sampling procedure, which will hopefully have a higher probability
of detecting outliers than simple random sampling.

Now the set of directions $\mathcal{D}$=$\left\{  \mathbf{u}_{k}%
,k\text{=}1,..,N_{\mathrm{kurt}},\ \mathbf{v}_{k},k\text{=}%
1,...,N_{\mathrm{SD}}\right\}  $ is used in the same manner as in the Stahel
(1981) -Donoho (1982) estimator to derive for each $\mathbf{z}_{i}$ an
outlyingness measure $t_{i}.$

In stage III the $t_{i}$s are used to make a preliminary classification of
outliers. Suspect observations are temporarily deleted, and the procedure is
repeated until no further changes occur.

The default values in the Matlab code kindly supplied by the authors are
$N_{\mathrm{Kurt}}$=2 and $N_{\mathrm{SD}}$=$10p.$

We now turn to fixing the problems shown in the former section. Experiments
showed that dealing with the case of dispersed outliers required increasing
$N_{\mathrm{SD}},$ and that a satisfactory choice was found to be
\begin{equation}
N_{\mathrm{SD}}=\max\left(  Mp,1000\right)  ~\mathrm{with~}M=50. \label{NSD}%
\end{equation}
Larger values of $M$ do not seem to yield an improvement.

This however did not fix the problem with concentrated outliers. Then the idea
was to use the $\mathbf{z}_{i}$'s themselves as directions, since if
$\mathbf{x}_{i}$ is an outlier, one would expect $\mathbf{z}_{i}$ to point in
the direction of $\mathbf{x}_{i}.$ Adding the set of $\mathbf{z}_{i}$'s to
$\mathcal{D}$ did in fact solve the problem. It may however become
computationally expensive for large $n$ since it is $O\left(  n^{2}\right)  .$
It was observed that contrary to expectation, the outliers corresponded to the
$\mathbf{z}_{i}$'s with \emph{smallest }norms. For this reason, rather than
adding to $\mathcal{D}$ the whole set of $n$ $\mathbf{z}^{\prime}$s, it was
decided to add the $m$ with smallest and the $m$ with largest norms, with
$m$=$\min\left(  5p,n/2\right)  .$

The next figure shows the results.%

\begin{center}
\includegraphics[
height=9.1752cm,
width=12.2033cm
]%
{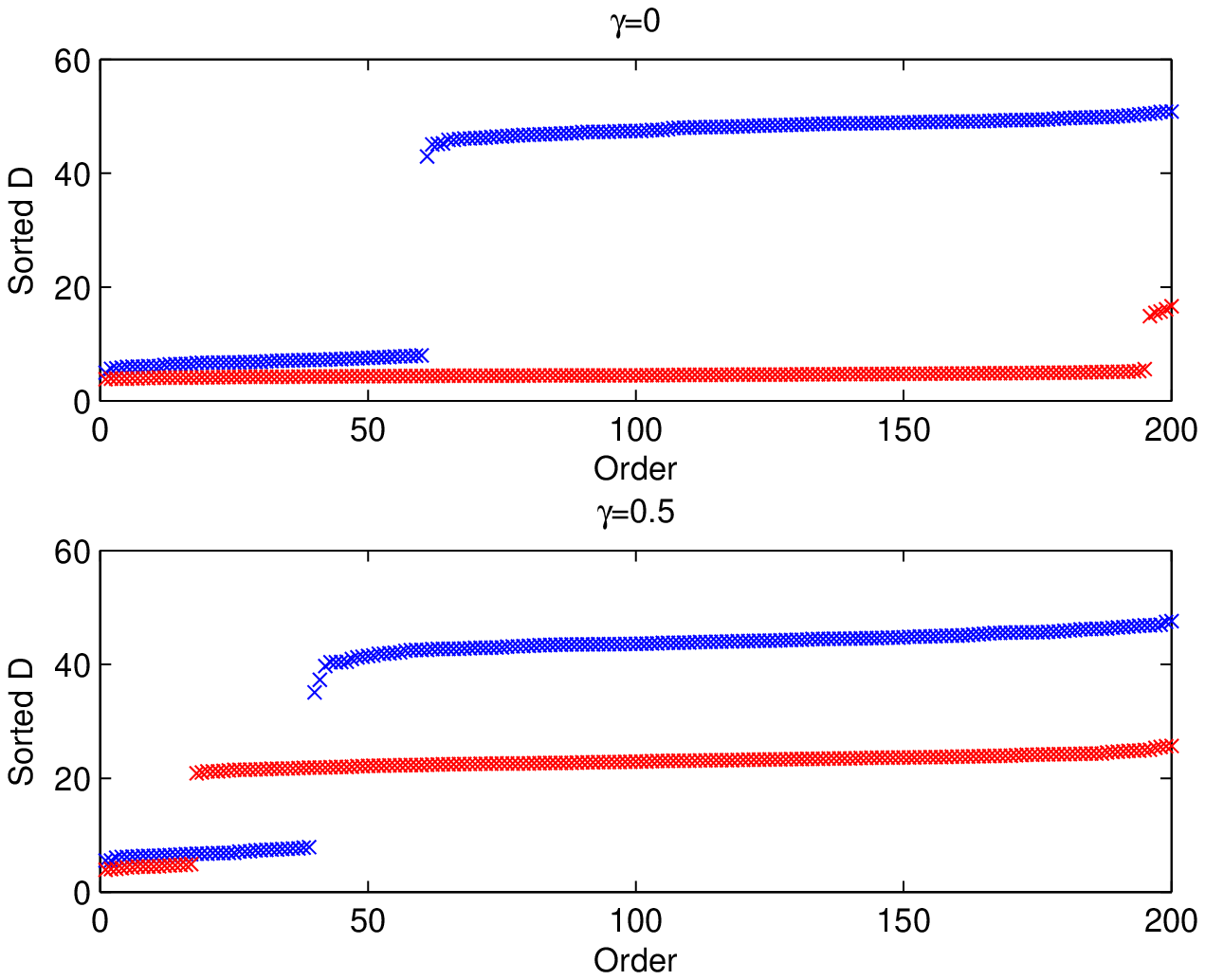}%
\\
Ordered values of $D\left(  \mathbf{\Sigma}\right)  $ of the Rocke estimator
for $K=13,$ starting from the old (blue) and new (red) versions of KSD.
\end{center}

It is seen that for $\gamma=0$ new version is very stable and yields much
lower values than the old one; for $\gamma=0.5$ it is more stable and its
values are generally much lower than those from the old one.

To have a more complete picture, the simulation was performed for
$K=1,2,...,30.$ The next figure shows the mean $D\left(  \widehat
{\mathbf{\Sigma}}\right)  $ of the Rocke estimator, corresponding to the old
and new versions of KSD.%

\begin{center}
\includegraphics[
height=9.1752cm,
width=12.2033cm
]%
{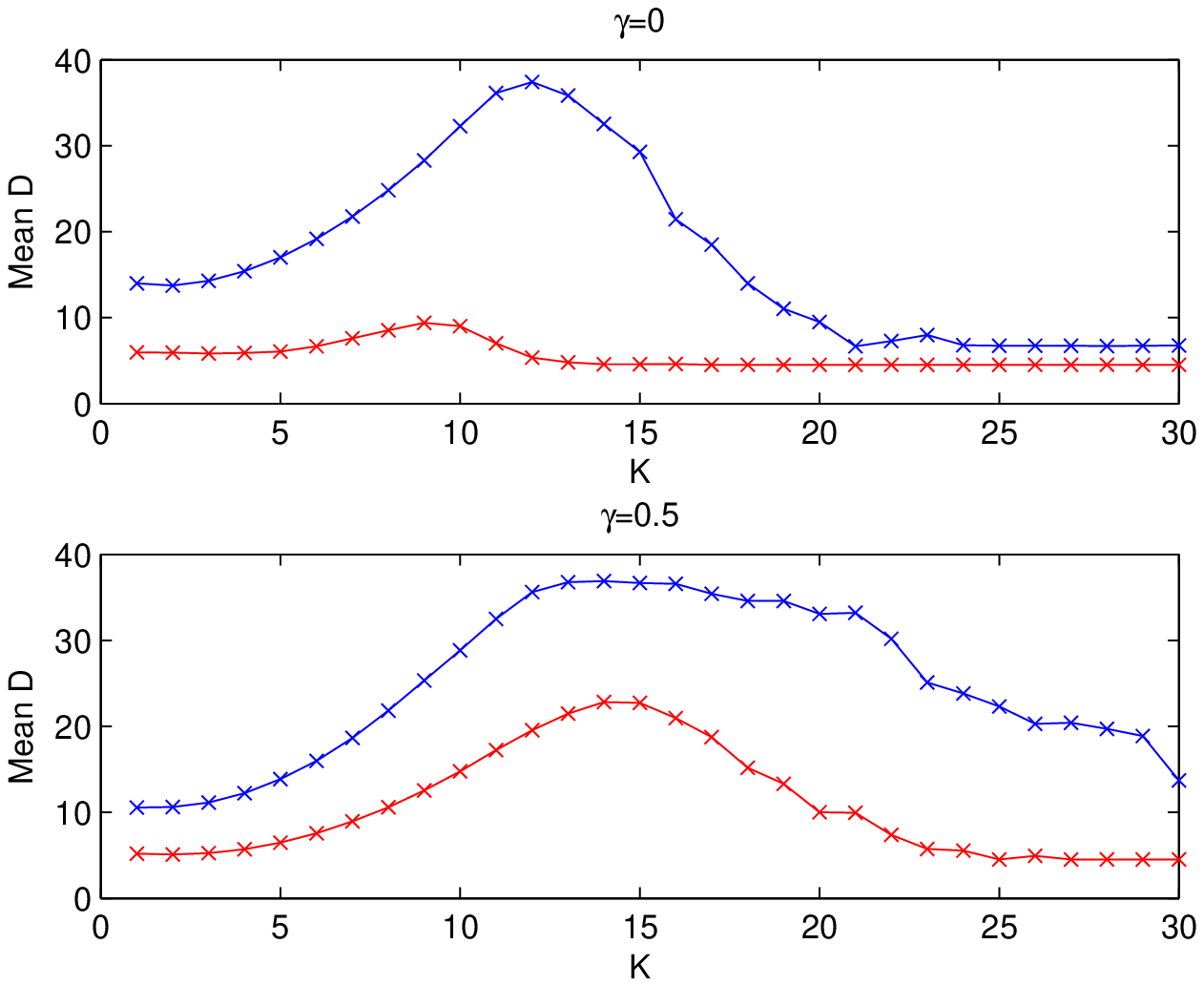}%
\\
Simulation for $p=30:$ mean $D\left(  \mathbf{\Sigma}\right)  $ starting from
the old (blue) and new (red) versions of KSD.
\end{center}

It is seen that the new version is a represents a clear improvement over the
old one. It may be argued that since for each $K$ the distribution of $D$ for
each estimator is bimodal, the mean is not a representative value. However,
using the median or a trimmed mean instead of the mean yields essentially the
same qualitative conclusions.

\section{Computing times}

The average computing times of Rocke's estimator based on the new to the old
version was computed for $p$ between 10 and 50. The following table gives the
results in seconds for the estimators computed in R on a PC with a 3.60 GHz
Intel Core processor with 16 GB RAM.

\begin{center}%
\begin{tabular}
[c]{rrrrr}\hline
$p$ & $n$ & New & Old & new/old\\\hline
20 & 100 & 0.030 & 0.008 & 3.750\\
& 400 & 0.036 & 0.015 & 2.400\\
50 & 250 & 0.134 & 0.056 & 2.393\\
& 1000 & 0.181 & 0.121 & 1.496\\
100 & 500 & 0.737 & 0.526 & 1.402\\
& 2000 & 0.970 & 0.871 & 1.113\\\hline
\end{tabular}

\end{center}

\bigskip It is seen that  the improvement does not have a high cost in terms
of computing performance. It is curious that the ratio new/old decreases with
$n$ and with $p.$ No explanation could be found for this fact.

\section{References}

Donoho, D. L. (1982). Breakdown properties of multivariate location
estimators,\ Ph. D. Qualifying paper, Harvard University.

Maronna, R.A. and Yohai, V.J. (2017). Robust and efficient estimation of high
dimensional scatter and location. \emph{Computational Statistics and Data
Analysis} (to appear). DOI 10.1016/j.csda.2016.11.006

Pe\~{n}a, D. and Prieto, F.J. (2007). Combining random and specific directions
for robust estimation of high-dimensional multivariate data. \emph{Journal of
Computational \& Graphical Statistics}, \textbf{16}, 228-254.

Rocke, D. (1996). Robustness properties of S-estimators of multivariate
location and shape in high dimension. \emph{The\ Annals of Statistics,}
\textbf{24}, 1327-1345.

Stahel, W.\ A. (1981).\ Breakdown of covariance estimators,\ Research report
31, Fachgruppe f\"{u}r Statistik, E.T.H. Z\"{u}rich.
\end{document}